\documentclass[%
reprint,
superscriptaddress,
showpacs,preprintnumbers,
 amsmath,amssymb,
 aps,
prd,
]{revtex4-1}

\usepackage{graphicx}
\usepackage{dcolumn}
\usepackage{bm}
\usepackage{bbold}
\usepackage{amssymb,amsmath}
\usepackage{hyperref}

\usepackage{color}
\usepackage{amsfonts}
\usepackage{subfigure}
\usepackage{array}

\newcolumntype{L}{>{\centering\arraybackslash}m{3cm}}

\definecolor{bjcol}{rgb}{1,.44,0.13}

\definecolor{blue}{rgb}{0,0,1}

\definecolor{green}{rgb}{0,1,0}

\definecolor{red}{rgb}{1,0,0}

\definecolor{gray}{rgb}{.5,.5,.5}

\definecolor{darkgreen}{rgb}{.0,.5,.0}

\def\Fig#1{Fig.~\ref{#1}}

\def\Eq#1{Eq.~(\ref{#1})}
\def\eq#1{(\ref{#1})}
\def\eqref#1{(\ref{#1})}

\def\lA0{{\langle A_0 \rangle}}
\def\bA0{{\bar{A}_0}}

\def\0#1#2{\frac{#1}{#2}}

\graphicspath{{./figures/}{./}}

\begin{document}

\preprint{}

\title{Four-fermion interactions and the chiral symmetry breaking in an external magnetic field
}

\author{Wei-jie Fu}
\email{wjfu@dlut.edu.cn}
\affiliation{School of Physics , Dalian University of Technology, Dalian, 116024,
  P.R. China}

\author{Yu-xin Liu}
\email[]{yxliu@pku.edu.cn}
\affiliation{Department of Physics and State Key Laboratory of Nuclear Physics and Technology, Peking University, Beijing 100871, China}
\affiliation{Collaborative Innovation Center of Quantum Matter, Beijing 100871, China}
\affiliation{Center for High Energy Physics, Peking University, Beijing 100871, China}


\begin{abstract}

We investigate the chiral symmetry and its spontaneous breaking at finite temperature and in an external magnetic field with four-fermion interactions of different channels. Quantum and thermal fluctuations are included within the functional renormalization group approach, and properties of the set of flow equations for different couplings, such as its fixed points, are discussed. It is found that external parameters, e.g. the temperature and the external magnetic field and so on, do not change the structure of the renormalization group flows for the couplings. The flow strength is found to be significantly dependent on the route and direction in the plane of couplings of different channels. Therefore, the critical temperature for the chiral phase transition shows a pronounced dependence on the direction as well. Given fixed initial ultraviolet couplings, the critical temperature increases with the increasing magnetic field, viz., the magnetic catalysis is observed with initial couplings fixed.

\end{abstract}

\maketitle


\section{\label{sec:intr}Introduction}

Recent studies on QCD and strongly interacting matter in extremely strong external magnetic fields have attracted lots of attentions, which are motivated, on one hand, by experimental observations of azimuthal charged-particle correlations in heavy-ion collisions at the Relativistic Heavy-Ion Collider (RHIC) and the LHC \cite{Abelev:2009ac,Abelev:2009ad,Adamczyk:2013hsi,Adamczyk:2014mzf,Abelev:2012pa}.  This phenomenon can be interpreted, although still being under discussion and recently challenged by relevant measurements in $p$-Pb collisions by the CMS Collaboration at LHC \cite{Khachatryan:2016got}, in the theoretical framework of the chiral magnetic effect (CME) \cite{Kharzeev:2004ey,Kharzeev:2007jp,Fukushima:2008xe}, whereof positive electric charges are separated from negative ones along the direction of the magnetic field produced in event-by-event noncentral heavy-ion collisions, due to the imbalanced chirality caused by a possible local violation of the parity symmetry. For more details about the CME and its recent progress, see e.g. \cite{Huang:2015oca} and references therein. 

On the other hand, how an extremely strong external magnetic field affects the spontaneous chiral symmetry breaking and the QCD chiral phase transition, catalyzing the symmetry breaking or inhibiting, is still in debate. Unlike many effective models which predict that, when the magnetic field strength $\bm{B}$ is enhanced, the (pseudo)-critical temperature $T_c$ for the chiral phase transition or crossover increases as well \cite{Fukushima:2010fe,Fraga:2013ova,Fu:2013ica,Zhang:2016qrl}, lattice QCD simulations found that $T_c$ decreases with increasing $\bm{B}$ \cite{Bali:2011qj,*Bali:2012zg}, which is because sea quarks coupled with the magnetic field increase the Polyakov loop, thus reduce the chiral condensate effectively \cite{Bruckmann:2013oba}. In the same time, based on other approaches, such as the Dyson-Schwinger equations, the magnetic catalysis is found for asymptotically large $\bm{B}$, while for intermediate $\bm{B}$, the inverse magnetic catalysis is observed, due to gluon screening effects and the strong coupling decreasing \cite{Mueller:2015fka}. This phenomenon is also called as the delayed magnetic catalysis, similarly found in a functional renormalization group (FRG) calculation as well \cite{Braun:2014fua}, in which the running of the four-fermion coupling is driven by not only the four-fermion but also the quark-gluon interactions. Besides gluonic effects as mentioned above (see e.g. \cite{Kojo:2012js,Mueller:2014tea} for more relevant discussions), the inverse catalysis can also be reproduced by including e.g., neutral meson effects \cite{Fukushima:2012kc,Mao:2016fha}, quark antiscreening \cite{Ferrer:2014qka}  or some other scenario \cite{Yu:2014xoa,Feng:2014bpa}.

Fukushima and Pawlowski have investigated the chiral symmetry breaking in an external magnetic field by studying the running four-fermion couplings within the FRG approach \cite{Fukushima:2012xw}, see e.g. \cite{Wetterich:1992yh,Pawlowski:2005xe,Pawlowski:2014aha,Pawlowski:2014zaa,Helmboldt:2014iya,Mitter:2014wpa,Braun:2014ata,Rennecke:2015eba,Fu:2015naa,Fu:2015amv,Wang:2015bky,Cyrol:2016tym,Fu:2016tey,Rennecke:2016tkm,Wang:2017vis} for more details about the FRG and its recent progresses in QCD. It is clearly shown there that how the magnetic field changes the pattern of the renormalization group (RG) flow for the four-fermion coupling and how the dimensional reduction takes place, in comparison to case without $\bm{B}$. Note also that only four-fermion interactions of scalar and pseudo-scalar channels, more specifically $\sigma$-$\pi$ channels, are included in analyses of \cite{Fukushima:2012xw}. Although scalar and pseudo-scalar channels play the most important role in the chiral symmetry breaking, they are not complete in the Dirac space, and are significantly influenced by other channels through quantum fluctuations, as we will show in what follows. Furthermore, four-fermion interactions with different channels are also indispensable to the dynamical hadronization technique \cite{Gies:2001nw,Gies:2002hq,Pawlowski:2005xe}, through which hadronic degrees of freedom emerge naturally as collective modes of quark-gluon dynamics in the low energy regime, and this technique has been successfully applied in recent rebosonized QCD computations \cite{Mitter:2014wpa,Braun:2014ata}. In this work we will perform the RG analyses for four-fermion interactions of different channels, investigate their mutual impacts on each other through quantum evolution, and study the influence on the chiral symmetry breaking and chiral phase transition in an external magnetic field.

The paper is organized as follows: In Sec.~\ref{sec:FRG} we investigate the four-fermion interactions of different channels within the FRG framework, and obtain a set of flow equations for the couplings. Then in Sec.~\ref{sec:num}  the structure of the flow equations is analyzed in detail. Flow diagram and relevant fixed points at vacuum are presented, and for the cases of finite temperature and an external magnetic field, numerical results are provided. A summary and conclusion can be found in Sec.~\ref{sec:sum}.


\section{\label{sec:FRG} Renormalization group flows for the four-fermion couplings}

We employ the following scale-dependent effective action for the two-flavor Nambu--Jona-Lasinio (NJL) model:
\begin{align}\nonumber 
  \Gamma_{k}=&\int_{x} \Big\{Z_{q,k}\bar{q}\gamma_{\mu}\partial_{\mu}q\, 
  + \frac{1}{2}\lambda_{-,k}\big[(\bar{q}\gamma_{\mu}q)^2-(\bar{q}i\gamma_{\mu}\gamma_{5}q)^2\big]\\[2ex]
  &+ \frac{1}{2}\lambda_{+,k}\big[(\bar{q}\gamma_{\mu}q)^2+(\bar{q}i\gamma_{\mu}\gamma_{5}q)^2\big]\Big\}\,,
\label{eq:action}\end{align}
with $\int_{x}=\int_0^{1/T}d x_0 \int d^3 x$ and the quark fields $q$, $\bar{q}$. $Z_{q,k}$ is the quark wave function renormalization; $\lambda_{-,k}$ and $\lambda_{+,k}$ are the four-fermion couplings for the $V-A$ and $V+A$ channels respectively, with $V$ and $A$ denoting vector and axial vector, where we have employed the notations in \cite{Braun:2011pp}. They are all scale-dependent with subscript $k$. Obviously, the four-fermion interactions in \Eq{eq:action} are $U(N_f)_{\mathrm{V}}\times U(N_f)_{\mathrm{A}}$ symmetric with flavor number $N_f=2$, and we don't take the $U(1)_{\mathrm{A}}$ breaking into account throughout this work. The $V+A$ channel is related to the scalar and pseudo-scalar ones through the Fierz transformations, which yield
\begin{align}
 &(\bar{q}\gamma_{\mu}q)^2+(\bar{q}i\gamma_{\mu}\gamma_{5}q)^2\nonumber \\[2ex]
 =&-\frac{4}{N_c}\Big\{\big[(\bar{q}\,T^0q)^2-(\bar{q}\,T^a\gamma_{5}q)^2\big]\nonumber \\[2ex]
 &+\big[(\bar{q}\,T^aq)^2-(\bar{q}\,T^0\gamma_{5}q)^2\big]\Big\}\nonumber \\[2ex]
 &-8\Big\{\big[(\bar{q}\,t^{\alpha}T^0q)^2-(\bar{q}\,t^{\alpha}T^a\gamma_{5}q)^2\big]\nonumber \\[2ex]
 &+\big[(\bar{q}\,t^{\alpha}T^aq)^2-(\bar{q}\,t^{\alpha}T^0\gamma_{5}q)^2\big]\Big\}\,,
\label{eq:fierzVpA}
\end{align}
with $T^{0}=\frac{1}{\sqrt{2N_{f}}}\mathbb{1}_{N_{f}\times N_{f}}$ and the $SU(N_{f})$ generators $T^a$ in flavor space, and the $SU(N_{c})$ generators $t^{\alpha}$ in color one. Those in the first set of square brackets on the r.h.s of \Eq{eq:fierzVpA} are the $\sigma$-$\pi$ channels, which are commonly used in the NJL model. Note that, besides the $\sigma$-$\pi$ channels, contributions also come from scalar triplets and the pseudo-scalar singlet, and even nontrivial colored channels with the $SU(N_{c})$ generators $t^{\alpha}$. Similarly, the $V-A$ channel, upon the implementation of the transformation, can be rewritten as 
\begin{align}
 &(\bar{q}\gamma_{\mu}q)^2-(\bar{q}i\gamma_{\mu}\gamma_{5}q)^2\nonumber \\[2ex]
 =&\frac{2}{N_c}\Big\{\big[(\bar{q}\,T^0\gamma_{\mu}q)^2-(\bar{q}\,T^ai\gamma_{\mu}\gamma_{5}q)^2\big]\nonumber \\[2ex]
 &+\big[(\bar{q}\,T^a\gamma_{\mu}q)^2-(\bar{q}\,T^0i\gamma_{\mu}\gamma_{5}q)^2\big]\Big\}\nonumber \\[2ex]
 &+4\Big\{\big[(\bar{q}\,t^{\alpha}T^0\gamma_{\mu}q)^2-(\bar{q}\,t^{\alpha}T^ai\gamma_{\mu}\gamma_{5}q)^2\big]\nonumber \\[2ex]
 &+\big[(\bar{q}\,t^{\alpha}T^a\gamma_{\mu}q)^2-(\bar{q}\,t^{\alpha}T^0i\gamma_{\mu}\gamma_{5}q)^2\big]\Big\}\,.
\label{eq:fierzVmA}
\end{align}
The $V-A$ channel, however, is invariant under the Fierz transformation, as only the Dirac structure is concerned. Therefore, it is orthogonal to the (pseudo)-scalar channels, while the $V+A$ channel is not. And since we are interested in the chiral symmetry and its breaking, does that mean the $V-A$ channel is not important and can be neglected? Obviously, it is not correct. We will show in what follows that quantum fluctuations in the $V-A$ channel affect those in the $V+A$ channel significantly, and thus the chiral symmetry breaking as well.

As the renormalization group (RG) scale $k$ in \Eq{eq:action} evolves from an ultraviolet (UV) cutoff scale $k=\Lambda$ down to the infrared $k=0$, quantum fluctuations with wavelength $\gtrsim 1/\Lambda$ are successively included in the effective action, through the Wetterich equation \cite{Wetterich:1992yh} as follows
\begin{align}
  \partial_{t}\Gamma_{k}&=\frac{1}{2}\mathrm{STr}\{\partial_{t} R_{k}(\Gamma_{k}^{(2)}+R_{k})^{-1}\}\nonumber \\[2ex]
 &=\frac{1}{2}\mathrm{STr}\{\tilde{\partial}_{t} \ln(\Gamma_{k}^{(2)}+R_{k})\}\,,
 \label{eq:WetterichEq}
\end{align}
with $t=\ln (k/\Lambda)$, where we have adopted the formalism in Ref.\cite{Gies:2001nw}; $\Gamma_{k}^{(2)}+R_{k}$, which is usually called as the fluctuation matrix, includes a regulator $R_k$ and the second derivative of the effective action with respect to all fields, i.e.,
\begin{align}
  (\Gamma_{k}^{(2)})_{ij}:=\frac{\overrightarrow{\delta}}{\delta\Phi_i}\Gamma_{k}\frac{\overleftarrow{\delta}}{\delta\Phi_j}\,,
 \label{eq:Gamma2}
\end{align}
with the super field $\Phi=(q,\bar q)$ for the NJL model. The super trace in \Eq{eq:WetterichEq} runs over momenta, fields and all other internal indices, and provides an additional minus sign for the fermionic part. Since in our case we only have fermionic quark fields, the minus sign is always there. The partial differentiation $\tilde{\partial}_{t}$ with a tilde in \Eq{eq:WetterichEq}  acts only on the regulator.

The fluctuation matrix can be rewritten as
\begin{align}
  \Gamma_{k}^{(2)}+R_{k}=\mathcal{P}+\mathcal{F}\,,
 \label{eq:flucmatrdecom}
\end{align}
where $\mathcal{P}$ is the matrix of inverse propagators with regulators, and $\mathcal{F}$ is the leftover part which includes the field dependence. Substituting \Eq{eq:flucmatrdecom} into \eq{eq:WetterichEq} and expanding $ \ln(\mathcal{P}+\mathcal{F})$ in order of $\mathcal{F}/\mathcal{P}$, one arrives at
\begin{align}
  \partial_{t}\Gamma_{k}=&\frac{1}{2}\mathrm{STr}\{\tilde{\partial}_{t} \ln(\mathcal{P}+\mathcal{F})\}=\frac{1}{2}\mathrm{STr}\tilde{\partial}_{t} \ln\mathcal{P}\nonumber \\[2ex]
 &+\frac{1}{2}\mathrm{STr}\tilde{\partial}_{t}\Big(\frac{1}{\mathcal{P}}\mathcal{F}\Big)-\frac{1}{4}\mathrm{STr}\tilde{\partial}_{t}\Big(\frac{1}{\mathcal{P}}\mathcal{F}\Big)^2+\cdots\,,
 \label{eq:WEqexpand}
\end{align}
from which we could obtain the flow equations for all the coupling in the effective action at appropriate expanding orders.

In this work we employ the $3d$ optimized regulator for the quark, i.e.,
\begin{align}
  R_{k}(q)&=Z_{q,k}i\vec{q}\cdot\vec{\gamma}\,r_F(\frac{{\vec{q}}^2}{k^2})\,,\quad \mathrm{with}\label{eq:regulator}\\[2ex]
  r_F(x)&=(\frac{1}{\sqrt{x}}-1)\Theta(1-x).
 \label{eq:rf}
\end{align}
Then through Eqs.~(\ref{eq:action}) (\ref{eq:Gamma2}) (\ref{eq:flucmatrdecom}) one has
\begin{align}
 \frac{1}{\mathcal{P}}&=\begin{pmatrix} 0 & S_k(q)\\[1ex]
                                                 -S_k(q)^T & 0
                                      \end{pmatrix}\,,
\label{}
\end{align}
with
\begin{align}
  S_k(q)=\frac{1}{Z_{q,k}i\big[q_0\gamma_0+(1+r_F)\vec{q}\cdot\vec{\gamma}\big]}\,,
 \label{}
\end{align}
and 
\begin{align}
 \mathcal{F}&=\begin{pmatrix} F_k^{qq} & F_k^{q\bar q}\\[1ex]
                                                F_k^{\bar qq} &  F_k^{\bar q\bar q}
                                      \end{pmatrix}\,,
\label{}
\end{align}
with
\begin{align}
  F_k^{qq}=&-(\lambda_{+,k}+\lambda_{-,k})(\bar q \gamma_{\mu})^T(\bar q \gamma_{\mu})\nonumber \\[1ex]
 &-(\lambda_{+,k}-\lambda_{-,k})(\bar q i\gamma_{\mu}\gamma_{5})^T(\bar q i\gamma_{\mu}\gamma_{5})\,,\\[2ex]
  F_k^{\bar q\bar q}=&-(\lambda_{+,k}+\lambda_{-,k})(\gamma_{\mu} q)(\gamma_{\mu} q)^T\nonumber \\[1ex]
 &-(\lambda_{+,k}-\lambda_{-,k})(i\gamma_{\mu}\gamma_{5} q)(i\gamma_{\mu}\gamma_{5} q)^T\,,\\[2ex]
  F_k^{\bar q q}=&(\lambda_{+,k}+\lambda_{-,k})\big[\gamma_{\mu} (\bar q\gamma_{\mu} q)+\gamma_{\mu} q \bar q\gamma_{\mu}\big]\nonumber \\[1ex]
 &+(\lambda_{+,k}-\lambda_{-,k})\big[(i\gamma_{\mu}\gamma_{5})(\bar q i\gamma_{\mu}\gamma_{5} q)\nonumber \\[1ex]
 &+i\gamma_{\mu}\gamma_{5}q \bar q\,i\gamma_{\mu}\gamma_{5}\big]\,,\\[2ex]
  F_k^{q \bar q}=&-(F_k^{\bar q q})^T\,.
 \label{}
\end{align}

%
\begin{figure}[t]
\includegraphics[width=0.45\textwidth]{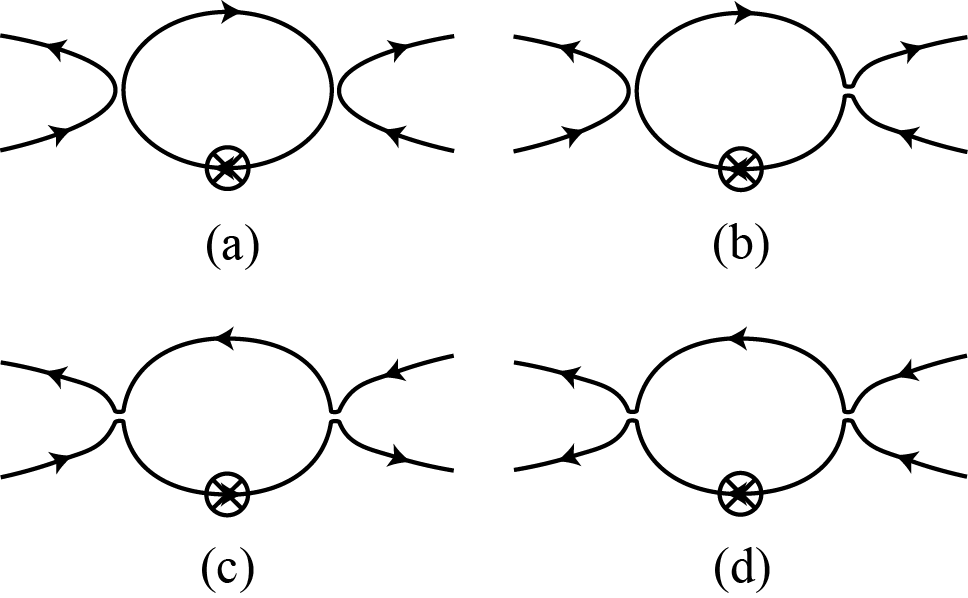}
\caption{Four different types of diagrams contributing to the flows of the four-fermion couplings, where prefactors for each diagram are not shown and the momentum dependence has been neglected. Note that the four-fermion vertices in these diagrams are denoted by two approaching lines, but not crossed, with each side associated with a Dirac index, e.g., $\gamma_{\mu}$ or $i\gamma_{\mu}\gamma_{5}$. The crossed circles stand for the regulator insertion.}\label{fig:fl4p}
\end{figure}
%

So far, we have all the elements to construct the flow equations for the four-fermion couplings. To begin,  we would like to mention that in this work the momentum dependence of the four-fermion couplings are neglected for simplicity, and the external momenta are assumed to be vanishing. It is found in \cite{Fukushima:2012xw} that the momentum dependence only has a minor quantitative effect. Inserting the expression of $\mathcal{F}/\mathcal{P}$ into \Eq{eq:WEqexpand} and only considering the second-order term, one finds that there are several different classes of diagrams contributing to the flow equations, which are shown in \Fig{fig:fl4p}. It is quite apparent that, among the four diagrams in \Fig{fig:fl4p}, only diagram (a) has a closed loop, which just corresponds to the usually called Hartree term, and is the leading-order term in the expansion of $1/(N_cN_f)$. In comparison to (a), there is a small leak on the right vertex in diagram (b) \footnote{It does not matter whether the leak appears on the right or left, because they can be related with each other through manipulations of symmetry, in the case that all external momenta are vanishing.}, and diagrams (c) and (d) have two leaks. Note that (c) and (d) are different, since their two connected fermionic lines are anti-parallel and parallel, respectively.

Inserting the regulator in \Eq{eq:regulator} into the diagrams in \Fig{fig:fl4p}, one can perform straightforward calculations for diagram (a). For others one needs additional Fierz transformations twice to obtain expressions, which have the same four-fermion interactions as \Eq{eq:action}. Then, the flow equations for the four-fermion couplings are readily obtained as
\begin{align}
  \partial_t \lambda_+&=-\big[2(N_cN_f+1)\lambda_+\lambda_-+3\lambda_+^2\big]\frac{l_2(k,T,\mu)}{k},\label{eq:flowlamp}\\[2ex]
  \partial_t \lambda_-&=-\big[(N_cN_f-1)\lambda_-^2+N_cN_f\lambda_+^2\big]\frac{l_2(k,T,\mu)}{k}\,.
 \label{eq:flowlamm}
\end{align}
We have verified that these equations are similar with those obtained in \cite{Braun:2011pp} for a fermionic model without color degrees of freedom. The threshold function is given by
\begin{align}
  l_2(k,T,\mu):=\frac{2\mathcal{F}_2(k,T,\mu)}{N_f}2N_f\int\frac{d^3q}{(2\pi)^3}\Theta(1-\frac{{\vec{q}}^2}{k^2})\,,
 \label{eq:l2}
\end{align}
with
\begin{align}
  \mathcal{F}_2(k,T,\mu)=&\frac{1}{4}\big[1-n_f(k,T,\mu)-n_f(k,T,-\mu)\big]\nonumber \\[1ex]
 &-\frac{k}{4T}\big[n_f(k,T,\mu)+n_f(k,T,-\mu)\nonumber \\[1ex]
 &-n_f^2(k,T,\mu)-n_f^2(k,T,-\mu)\big]\,,
 \label{}
\end{align}
and the fermionic distribution function
\begin{align}
  n_f(k,T,\mu)=&\frac{1}{e^{(k-\mu)/T}+1}\,.
 \label{}
\end{align}

To be proceeded, the flow equations can be easily extended to the case of a finite external magnetic field. Assuming a spatially homogeneous, temporally independent  magnetic field $\bm{B}$ aligning along the $z$ axis, due to the quantization of the transverse momenta into Landau levels, one has
\begin{align}
  \vec{q}^2=q_z^2+2\lvert q_f eB\rvert n\,,
 \label{}
\end{align}
with the electric charge $q_f e$. And the 3-momentum integral in \Eq{eq:l2} is modified to 
\begin{align}
  &2N_f\int\frac{d^3q}{(2\pi)^3}\Theta(1-\frac{{\vec{q}}^2}{k^2})\nonumber \\[1ex]
 \longrightarrow&\frac{1}{2\pi^2}\sum_{f=u,d}\lvert q_f eB\rvert\sum_{n=0}^{N_{k,f}}\alpha_n\sqrt{k^2-2\lvert q_f eB\rvert n}\,,
 \label{}
\end{align}
with $\alpha_0=1$ for the lowest-order Landau level and $\alpha_{n>0}=2$; $N_{k,f}$ is given by
\begin{align}
  N_{k,f}=\theta_g\left(\frac{k^2}{2\lvert q_f eB\rvert}\right)\,,
 \label{}
\end{align}
with $\theta_g(n+x)=n$ for $0<x<1$ and integer $n$.

Therefore, if there is a finite magnetic field, flow equations in (\ref{eq:flowlamp}) and (\ref{eq:flowlamm}) are not changed, but with the threshold function replaced with
\begin{align}
  l_2(k,T,\mu, eB)=&\frac{\mathcal{F}_2(k,T,\mu)}{\pi^2 N_f}\sum_{f=u,d}\lvert q_f eB\rvert\nonumber \\[1ex]
 &\times\sum_{n=0}^{N_{k,f}}\alpha_n\sqrt{k^2-2\lvert q_f eB\rvert n}\,,
 \label{eq:l2}
\end{align}

\section{\label{sec:num}Numerical results}

%
\begin{figure}[t]
\includegraphics[width=0.53\textwidth]{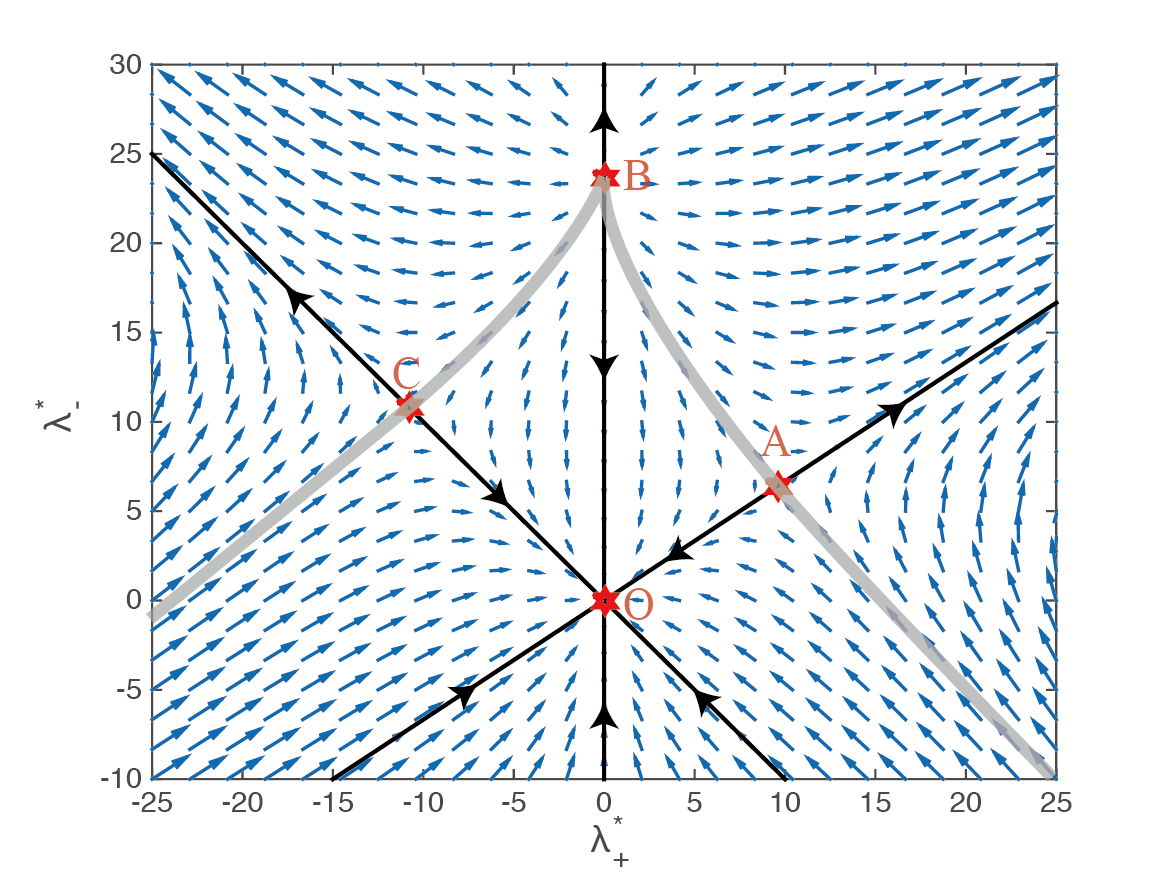}
\caption{Diagram of the infrared flow for the four-fermion couplings at vacuum, which is shown by the vector fields ($\beta_+$, $\beta_-$) in Eqs.~(\ref{eq:betaplus})~(\ref{eq:betaminus}) in the plane of $\lambda_+^*$ and $\lambda_-^*$. The fixed points in the plane are labelled by the red stars, and the black solid lines with arrows denote three specific flow lines, see text for more details.}\label{fig:flowfield}
\end{figure}
%

It is interesting to note that, from the flow equations for the four-fermion couplings in Eqs.~(\ref{eq:flowlamp})~(\ref{eq:flowlamm}), all the external influences, such as the temperature, chemical potential, and the magnetic field are implemented only through the threshold function $l_2$. Thus the external parameters do not change the structure of the flow equations. For the case of vacuum, $l_2(k)=k^3/(6\pi^2)$, it is more convenient to introduce the dimensionless couplings, i.e.,
\begin{align}
  \lambda_+^*=k^2\lambda_+\quad\mathrm{and}\quad \lambda_-^*=k^2\lambda_-\,,
 \label{}
\end{align}
whose flow equations are readily obtained as
\begin{align}
  \partial_t \lambda_+^*=-\beta_+\quad\mathrm{and}\quad \partial_t \lambda_-^*=-\beta_-\,,
 \label{eq:dtlambstar}
\end{align}
with
\begin{align}
  \beta_+=&\frac{1}{6\pi^2}\big[2(N_cN_f+1)\lambda_+^*\lambda_-^*+3(\lambda_+^*)^2\big]-2\lambda_+^*\,,\label{eq:betaplus}\\[1ex]
\beta_-=&\frac{1}{6\pi^2}\big[(N_cN_f-1)(\lambda_-^*)^2+N_cN_f(\lambda_+^*)^2\big]-2\lambda_-^*\,.
 \label{eq:betaminus}
\end{align}
The vector field ($\beta_+$, $\beta_-$), flowing toward the infrared (IR) in the plane of $\lambda_+^*$ and $\lambda_-^*$, is plotted in \Fig{fig:flowfield}. Apparently, there are four fixed points in the flow diagram, which are labelled by red stars. Of these points, the one at the origin, i.e. point $O$ with $\lambda_+^*=\lambda_-^*=0$, is the IR fixed point, while the other three belong to the kind of UV ones, and their coordinates ($\lambda_+^*$, $\lambda_-^*$) are given by
\begin{align}
  A=&\left(\frac{12(3+N_cN_f)\pi^2}{9+5N_cN_f+2N_c^2N_f^2},\;\frac{12 N_cN_f\pi^2}{9+5N_cN_f+2N_c^2N_f^2}\right),\nonumber\\[1ex]
  B=&\left(0,\;\frac{12 \pi^2}{N_cN_f-1}\right),\nonumber\\[1ex]
  C=&\left(-\frac{12 \pi^2}{2N_cN_f-1},\;\frac{12 \pi^2}{2N_cN_f-1}\right).
 \label{}
\end{align}
Drawing a straight line across both the IR fixed point $O$ and any one of the UV points, say point $A$, one obtains a specific line of renormalization flow $OA$. A system, initially located on the line, won't flow away from it forever. This is quite obvious for line $OB$ from Eqs.~(\ref{eq:flowlamp})~(\ref{eq:flowlamm}) with $\lambda_+=0$. It can be easily checked that this also holds for lines $OA$ and $OC$. For $OA$, performing the following rotation:
\begin{align}
 \begin{pmatrix} \lambda_+^{\prime} \\[1ex] \lambda_-^{\prime} \end{pmatrix}&=\begin{pmatrix} \cos\theta & \sin\theta\\[1ex] -\sin\theta &  \cos\theta \end{pmatrix} \begin{pmatrix} \lambda_+ \\[1ex] \lambda_- \end{pmatrix}\,,
\label{}
\end{align}
with
\begin{align}
  \cos\theta=&\frac{N_cN_f+3}{\sqrt{(N_cN_f)^2+(N_cN_f+3)^2}},\nonumber\\[1ex]
  \sin\theta=&\frac{N_cN_f}{\sqrt{(N_cN_f)^2+(N_cN_f+3)^2}}.
 \label{}
\end{align}
one arrives at
\begin{align}
  \partial_t \lambda_+^{\prime}=&-\big[(2N_c^2N_f^2+5N_cN_f+9){\lambda_+^{\prime}}^2+2(N_cN_f+3)\nonumber\\[1ex]
&\times\lambda_+^{\prime}\lambda_-^{\prime}-3N_cN_f{\lambda_-^{\prime}}^2\big]\frac{l_2}{k}\nonumber\\[1ex]
&\times\frac{1}{\sqrt{(N_cN_f)^2+(N_cN_f+3)^2}},\label{}\\[2ex]
  \partial_t \lambda_-^{\prime}=&-\lambda_-^{\prime}\big[-4N_cN_f\lambda_+^{\prime}+(2N_c^2N_f^2+2N_cN_f-3)\lambda_-^{\prime}\big]\nonumber\\[1ex]
&\times\frac{l_2}{k}\frac{1}{\sqrt{(N_cN_f)^2+(N_cN_f+3)^2}}\,.
 \label{}
\end{align}
line $OA$ just corresponds to the solutions in the equations above with $\lambda_-^{\prime}=0$, and the two equations are reduced to a single one as follows
\begin{align}
  \partial_t \lambda_+^{\prime}=&-c\frac{l_2}{k}{\lambda_+^{\prime}}^2\,,
 \label{eq:lambppri}
\end{align}
with
\begin{align}
  c=&\frac{(2N_c^2N_f^2+5N_cN_f+9)}{\sqrt{(N_cN_f)^2+(N_cN_f+3)^2}}\simeq 10.3\,.
 \label{}
\end{align}
In the same way, this property also applies to line $OC$. 

%
\begin{figure}[t]
\includegraphics[width=0.53\textwidth]{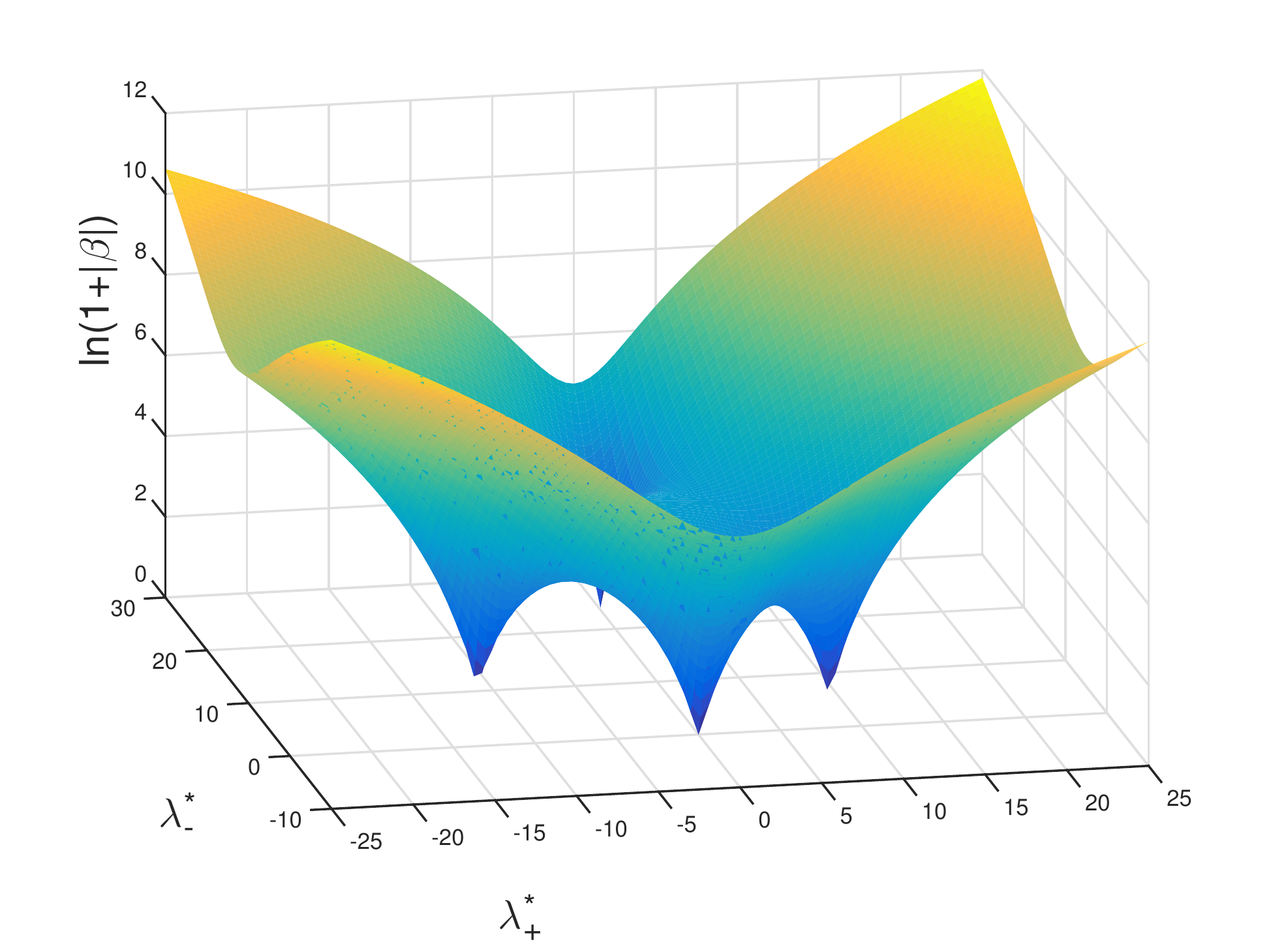}
\caption{Magnitude of the flow vector ($\beta_+$, $\beta_-$) in Eqs.~(\ref{eq:betaplus})~(\ref{eq:betaminus}), plotted in $\ln(1+|\beta|)$ as a function of $\lambda_+^*$ and $\lambda_-^*$.}\label{fig:flowmod}
\end{figure}
%

In summary, the two entangled flow equations for the $V-A$ and $V+A$ couplings are decomposed into a simple single flow equation, when the system is located on a straight line which connects the IR and UV fixed points. Properties of the flow equation on these lines and its solutions, at vacuum and at finite $T$ or/and $\bm{B}$, have been studied in detail in \cite{Fukushima:2012xw}, and all results there can be directly applied to the reduced flow equation in this work. Taking \Eq{eq:lambppri} for example, only when the coupling strength $\lambda_{\Lambda+}^{\prime}$ at the initial UV evolution scale $\Lambda$ at vacuum fulfills 
\begin{align}
  \Lambda^2\lambda_{\Lambda+}^{\prime}&\geq \frac{12\pi^2}{c}\,,
 \label{eq:chiralbreaking}
\end{align}
one has the spontaneous chiral symmetry breaking. As shown in \Fig{fig:flowfield}, line $OA$ is divided into two parts by point $A$, and the condition in \Eq{eq:chiralbreaking} just corresponds to the case that, at $k=\Lambda$, the system is located at point $A$ or on the higher part of line $OA$, which flows away from the IR fixed point $O$. Apparently, systems on the other sector of line $OA$ all flow toward the IR fixed point $O$ and there are no spontaneous chiral symmetry breaking. Similar analysis is also applicable to lines $OB$ and $OC$. It is straightforward to extend our analysis to the whole plane of $\lambda_+^*$ and $\lambda_-^*$. We have depicted schematically two gray lines crossing the three UV fixed points in \Fig{fig:flowfield}, separating the plane into two regions, of which the above one is that of spontaneous chiral symmetry breaking, while that below has no chiral symmetry breaking. One should note that the strength of flow is dependent on the route where it goes. As for the three specific lines $OA$, $OB$ and $OC$, they have a different coefficient $c$ in their respective flow equations as in \Eq{eq:lambppri}; furthermore, the dependence of the flow strength on routes can also be found in \Fig{fig:flowmod}, where the magnitude of the vector ($\beta_+$, $\beta_-$) in Eqs.~(\ref{eq:betaplus})~(\ref{eq:betaminus}), i.e., $|\beta|=\sqrt{\beta_+^2+\beta_-^2}$ is shown in the plane of $\lambda_+^*$ and $\lambda_-^*$. Here we have used $\ln(1+|\beta|)$ instead of $|\beta|$ directly for the convenience of plotting. A pronounced dependence of the flow strength on the route and direction is observed in this figure.

Moreover, the spontaneously broken chiral symmetry at vacuum can be restored at finite $T$, and the corresponding critical temperature is known from \cite{Fukushima:2012xw} as
\begin{align}
  T_c=\left(\frac{\Lambda^2}{\pi^2}-\frac{12}{\lambda_{\Lambda+}^{\prime}c}\right)^{\frac{1}{2}}
 \label{}\,,
\end{align}
with line $OA$ for instance. Although one could not obtain a set of equations similar with Eqs.~(\ref{eq:dtlambstar}) through (\ref{eq:betaminus}) at finite $T$, it can be still regarded that the three UV fixed points in \Fig{fig:flowfield} move effectively along the three straight line respectively, all in directions away from the IR fixed point $O$.

When the external magnetic field $\bm{B}$ is nonzero and $T=0$, It is found in \cite{Fukushima:2012xw} that there is always a chiral symmetry breaking in the lowest-Landau level approximation, no matter how small the coupling is, because of the dimensional reduction. This is the magnetic catalysis of the chiral symmetry breaking. This conclusion for a fermionic system with one coupling is still valid in our case, and the three UV fixed points in \Fig{fig:flowfield} move toward, and coincide at the origin $O$ point, when $\bm{B}\neq 0$ and $T=0$ in the lowest-Landau level approximation. In the same way, when $\bm{B}\neq 0$, $T\neq0$ and beyond the lowest-Landau level approximation, the three UV fixed points $A$, $B$ and $C$ are not at the origin point $O$ any more, but they all move toward $O$ with the increase of the magnetic field, indicating that the region of chiral symmetry breaking above the gray line in \Fig{fig:flowfield} is enlarged, which is also an appearance of the magnetic catalysis.
 
%
\begin{figure}[t]
\includegraphics[width=0.48\textwidth]{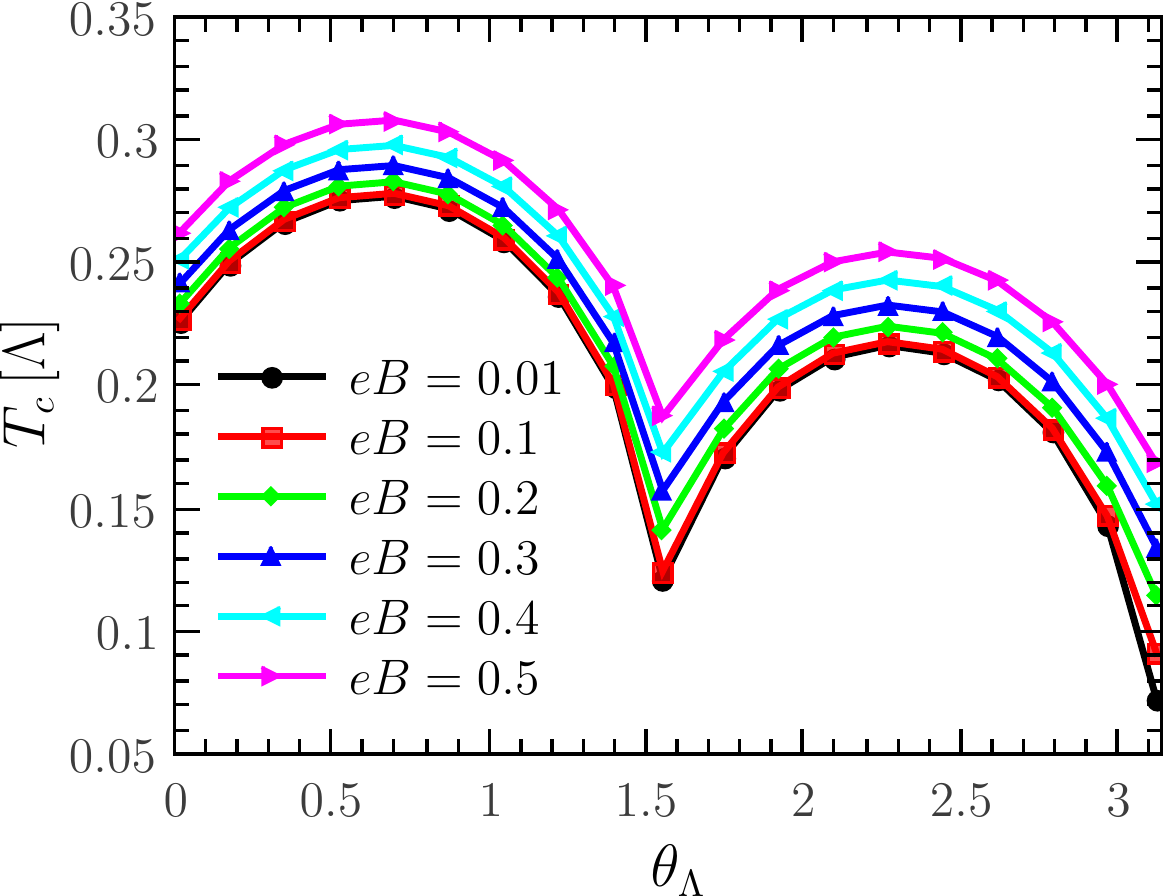}
\caption{Critical temperature for the chiral phase transition $T_c$ in unit of the initial UV renormalization group scale $\Lambda$, as a function of the angle $\theta_{\Lambda}$ with $\lambda_{\Lambda}\Lambda^2=25$. Quantities $\theta_{\Lambda}$ and $\lambda_{\Lambda}$ are defined in \Eq{eq:thetalam}. Calculations for several values of the magnetic field $eB$ in unit of $\Lambda^2$ are compared.}\label{fig:Tc-theta}
\end{figure}
%

Given a set of four-fermion couplings at the initial UV evolution scale $\Lambda$, i.e., $\lambda_{\Lambda+}$ and $\lambda_{\Lambda-}$, the flow equations in (\ref{eq:flowlamp}) and (\ref{eq:flowlamm}) can be solved numerically. In order to investigate the dependence of the flow on the route and direction, we parametrize $\lambda_{\Lambda+}$ and $\lambda_{\Lambda-}$ as follow
\begin{align}
  \lambda_{\Lambda+}=\lambda_{\Lambda}\cos \theta_{\Lambda},\quad \lambda_{\Lambda-}=\lambda_{\Lambda}\sin \theta_{\Lambda}\,.
 \label{eq:thetalam}
\end{align}
In \Fig{fig:Tc-theta} we show the dependence of $T_c$, the critical temperature for the chiral phase transition, on the direction in the plane as depicted in \Fig{fig:flowfield}, which is represented by the angle $\theta_{\Lambda}$ defined above. Here we choose $\lambda_{\Lambda}\Lambda^2=25$ which guarantees that 
there is always a chiral symmetry breaking with $\theta_{\Lambda}\in [0\,,\pi]$, as shown apparently in \Fig{fig:flowfield}. Furthermore, calculated results for several different values of the magnetic field strength are compared, and the chemical potential is assumed to be vanishing. One can easily find that curves in \Fig{fig:Tc-theta} have a quite obvious feature that the shape of these curves looks like a deformed letter ``M''. The two directions relevant  to the two peaks in \Fig{fig:Tc-theta} are almost collinear to lines $OA$ and $OC$ in \Fig{fig:flowfield}, respectively. This is reasonable and can be anticipated, because one can observe in \Fig{fig:flowmod} that the flow strength obtains its maximum in these two directions. Comparing $OA$ and $OC$, we can find that $OA$ is the direction, which breaks the chiral symmetry most effectively. In contrast, values of the critical temperature are relatively small in the directions with $\theta_{\Lambda}=0$, $\pi/2$ and $\pi$. Moreover, the dependence of the critical temperature on the magnetic field strength is also investigated in \Fig{fig:Tc-theta}, and we find that $T_c$ increases with the increasing $eB$, which is a feature of the magnetic catalysis of the spontaneous chiral symmetry breaking.


\section{\label{sec:sum} Summary and conclusions}

In this work we have studied the chiral symmetry and its spontaneous breaking at finite temperature and in an extremely strong external magnetic field, in a model with four-fermion interactions of $V+A$ and $V-A$ channels, with $V$ and $A$ denoting vector and axial vector respectively. The $V+A$ channel is related to the usually employed $\sigma$-$\pi$ one through the Fierz transformation, while $V-A$ channel is orthogonal to $V+A$. Quantum and thermal fluctuations are encoded within the functional renormalization group approach, and a set of flow equations for the couplings of $V+A$ and $V-A$ channels, i.e., Eqs. (\ref{eq:flowlamp}) and (\ref{eq:flowlamm}) is obtained.

There are one infrared and three ultraviolet fixed points in the flow diagram in terms of dimensionless couplings at vacuum. Three straight flow lines connect the three UV fixed points with the IR one, respectively. Inclusion of external parameters, such as the temperature and the magnetic field and so on, only moves the three UV fixed points on their respective flow lines, while the directions of the three straight flow lines are not changed. This is because external parameters enter into flow equations (\ref{eq:flowlamp}) and (\ref{eq:flowlamm}) only through the threshold function $l_2$, and the ratio between Eq. (\ref{eq:flowlamp}) and (\ref{eq:flowlamm}) is independent of $l_2$. In another word, external parameters do not modify the structure of the flow equations. 

Furthermore, we have found that the flow strength is significantly dependent on the route and direction in the plane of couplings of different channels, which results in that the critical temperature $T_c$ for the chiral phase transition has a pronounced dependence on the direction as shown in \Fig{fig:Tc-theta}. In the specific model used in this work, $OA$ in \Fig{fig:flowfield} is found to be the direction, which breaks the chiral symmetry most effectively.

We also find that the magnetic catalysis effect, i.e., the critical temperature increases with the increasing magnetic field. Note, however, that this conclusion is based on the implicit assumption that all the initial UV couplings are fixed when the magnetic field strength is increased. This is obviously not true, once other dynamical degrees of freedom, such as the gluonic fields, are included, as done in \cite{Mueller:2015fka,Braun:2014fua}. Considering the quite obvious dependence of the flow strength on the direction and route, it is very interesting to extend our work to include the gluonic fluctuations, which will give more information about the magnetic catalysis or inverse magnetic catalysis. Furthermore, we noticed interestedly that J. Braun {\it et al.} have found that the inclusion of Fierz-complete four-fermion interactions has a significant impact on the QCD phase structure \cite{Braun:2017srn}, when we are very close to the accomplishment of this work.

\begin{acknowledgments}

The work was supported by the National Natural Science Foundation of China under Contracts No. 11435001; the National Key Basic Research Program of China under Contract Nos. G2013CB834400
 and 2015CB856900; the Fundamental Research Funds for the Central Universities under Contract No. DUT16RC(3)093.

\end{acknowledgments}


\bibliography{ref-lib.bib}

\end{document}